\documentstyle[aps,prl,twocolumn]{revtex}

\def\be{\begin{equation}}
\def\ee{\end{equation}}
\def\bea{\begin{eqnarray}}
\def\eea{\end{eqnarray}}
\def\bma{\begin{mathletters}}
\def\ema{\end{mathletters}}

\def\0{\overline{0}}

\def\q0{\underline{0}}

\def\U{{ U}}

 \def\ket#1{|#1\rangle}

\begin{document}
\draft

\title{Time--optimal Hamiltonian simulation and gate synthesis \\ using homogeneous local unitaries}

\author{Ll. Masanes$^1$, G. Vidal$^{2,3}$ and J. I. Latorre$^1$}

\address{
$^{1}$Dept. d'Estructura i Constituents de la Mat\`eria, Univ. Barcelona, 08028. Barcelona, Spain.
\\
$^{2}$Institut f\"ur Theoretische Physik, Universit\"at Innsbruck, A-6020 Innsbruck, Austria.\\
$^{3}$ Institute for Quantum Information, California Institute of Technology, Pasadena, CA 91125, USA.}

\date{\today}

\maketitle

\begin{abstract}
Motivated by experimental limitations commonly met in the design of solid state quantum computers,
we study the problems of non--local Hamiltonian simulation and non--local gate synthesis when only {\em homogeneous} local unitaries are performed in order to tailor the available interaction. Homogeneous (i.e. identical for all subsystems) local manipulation implies a more refined classification of interaction Hamiltonians than the inhomogeneous case, as well as the loss of universality in Hamiltonian simulation. For the case of symmetric two--qubit interactions, we provide time--optimal protocols for both Hamiltonian simulation and gate synthesis. 
\end{abstract}


\section{\label{sec1}Introduction}

\subsection{Simulation of non--local Hamiltonians and synthesis of non--local gates}

%
%

The simulation of Hamiltonians has attracted consi\-derable attention in several areas of quan\-tum physics, including those of nuclear magnetic resonance \cite{NMR} (NMR), quantum control \cite{control} and quantum information \cite{Duer,Dodd,Wocjan,IBM,Bremner}. Let us consider a quantum system that naturally evolves according to some Hamiltonian $H$ for a time $t$. By interspersing its evolution $\exp(-iHt)$ with a sequence of fast unitary operations $W_k$ \cite{foot1}, another evolution 
\be
e^{-iH't'} = e^{-iHt_n}W_n\cdots e^{-iHt_2}W_2 e^{-iHt_1}W_1
\label{simevol}
\ee
is obtained which, when $|Ht_k|<< 1$, approximately corresponds to having the Hamiltonian 
\be
H' = \frac{c}{t} \sum_k t_k V_k H V_k^{\dagger},
\label{simham}
\ee
$V_k^{\dagger}\equiv W_k\cdots W_1$ [we assume $V_n= I$], acting for a time $t'\equiv t/c$, where $c\geq 0$ is 
the time overhead of the simulation.
This means that we can effectively generate, given an available Hamiltonian $H$ and some control operations $W_k$, a new Hamiltonian $H'$, which may otherwise not be accessible experimentally. Hamiltonian simulation is therefore a powerful technique to control the evolution of a quantum system. As such, it has a broad range of potential applications, and may eventually become commonplace in experimental implementations of quantum information and quantum computation.

%
%

The synthesis of a gate $U\equiv \exp(-iH'T)$ by using $H$ and control operations $W_k$ is a closely related issue, for $U$ can be achieved by simulating $H'$ for a time $T$. We notice, however, that inequivalent sequences of control operations $W_k$ may also lead to the same gate $U$ ---the only requirement is that at the end of the protocol the gate has been performed--- and thus simulating $H'$ may not be the most convenient way to proceed.

A series of recent contributions \cite{Duer,Dodd,Wocjan,IBM,Bremner} has recently addressed, in the context of multipartite quantum systems, the problem of simulating one non--local Hamiltonian (or interaction) by using another non--local Hamiltonian and local unitary transformations (LU) as control operations. In particular, special emphasis has been placed on assessing the universality of this model \cite{Dodd,IBM,Bremner}, as well as on determining time-optimal simulation protocols and developing a quantitative theory of interactions based on their simulation capabilities \cite{Duer,IBM}. Similarly, time-optimal protocols for non--local gate synthesis have been thoroughly analyzed in the context of quantum control \cite{InteractionCost2}, where the problem has been  reduced, in the case of a two-qubit system, to a problem of non--local Hamiltonian
 simulation. In \cite{InteractionCost} the corresponding optimal non--local Hamiltonian, and thereby the minimal time for two--qubit gate synthesis, have been presented.

\subsection{Hamiltonian simulation in solid state physics: \\homogeneous local manipulation}

%
%

In most of previous work on non-local Hamiltonian simulation, arbitrary LU are assumed to be available as control operations. This model is motivated by the fact that in a number of experimental schemes for quantum information processing (using e.g. ion traps, neutral atoms or photons \cite{Fort}) independent LU operations can be performed on each of the interacting systems supporting the qubits. 
In solid state implementations, however, the relevant systems are frequently too close to each other to be addressed individually.
A good example of this is a set of spins in a NMR molecule or in a lattice of quantum dots, where an extraordinarily focused (and thus unfeasible) magnetic field would be required to independently address each spin. Therefore, and due to experimental limitations, the current model for interaction simulation does not apply to a number of solid state schemes. 

Most relatedly, an alternative, solid-state oriented model for quantum computation has recently been put forward \cite{Di}. It overcomes the above lack of controllability by encoding logical qubits in convenient subspaces of few physical qubits, in a way that universal computation can be accomplished by just tunning the intensity of a specific non--local Hamiltonian ---such as the isotropic exchange interaction, naturally available in many systems.
 Notice, however, that in solid state not all forms of local manipulation are equally inaccessible. In particular, a uniform magnetic field can be easily applied on all spins, inducing the same local evolution on each of them and thereby producing what we shall call an homogeneous local unitary transformation (HLU). In this sense, HLU appear as a rather restricted, but definitely feasible, set of control operations 
in solid state physics, that may be worth considering.

%
%

\subsection{Results}

In this paper we address the problems of non--local Hamiltonian simulation and gate synthesis using a given non--local Hamiltonian and fast HLU. Our results, mostly concerned with two-body interactions in multi-qubit systems, imply that an interaction, when only supplemented with HLU, does not constitute a universal resource for simulating other interactions or synthesizing arbitrary gates, in sharp contrast with the inhomogeneous LU model \cite{Dodd,IBM,Bremner}. Nevertheless, the HLU model still allows for a rich variety of Hamiltonian simulations and gate synthesis. In particular, HLU can be used to eliminate unwished anisotropic terms (as coming e.g. from a spin-orbit interaction) in a system of spins with imperfect isotropic exchange interaction. This task is of practical relevance, for instance, in the context of encoded universality, and has been considered in \cite{Di2}. Therefore the HLU model for non-local Hamiltonian simulation provides us with a complementary or alternative technique to that discussed in \cite{Di2}.

We shall present the following results.

\vspace{2mm}

\noindent ($i$) Necessary and sufficient conditions for two-qubit Hamiltonian simulation under HLU, both for symmetric and antisymmetric interactions. These results also apply to some multiple-qubit scenarios with two-body interactions.

\vspace{2mm}

\noindent ($ii$) Time-optimal synthesis of two-qubit gates using a symmetric interaction.

\vspace{2mm}

\noindent ($iii$) Recovery of universality in gate synthesis using HLU and just one initial and one final non-homogeneous LU per two-qubit gate.

\vspace{2mm}

A remarkable feature of the HLU model of simulation will be that it is ruled by the usual majorization relation \cite{Bhatia}, instead of the special majorization relation that rules the non--homogeneous LU model \cite{IBM}. As we shall explain, this gives a neat picture of which two-qubit Hamiltonian simulations are possible in the symmetric case: {\em $H$ can simulate $H'$ if and only if this implies a gain of isotropy in the interaction.}

The organization of this paper goes as follows. In sect. II we address the problem of two-qubit Hamiltonian simulation and present the optimal solution in the case of symmetric and antisymmetric interactions. Using these results, the time-optimal synthesis of symmetric gates using a symmetric interaction is analized in sect. III. In sect. IV we discuss the synthesis of symmetric gates using an arbitrary interaction. In sect. V we also show how to achieve any two-qubit gate by allowing for two non--homogeneous interventions. The paper finishes with some conclusions.

\section{\label{sec2}Optimal Hamiltonian simulation using HLU}

 Consider a collection of $N$ interacting systems, say $N$ qubits for concreteness, that evolve according to a Hamiltonian $H_{12\cdots N}$. If only fast HLU can be performed on the qubits in order to control their evolution, i.e. $W_k = w_k^{\otimes N}$ in Eq. (\ref{simevol}), then, according to Eq. (\ref{simham}), any simulated Hamiltonian $H_{12\cdots N}'$ must be of the form
\be
H_{12\cdots N}' = c\sum_k p_k (v_k^{\otimes N}) H_{12\cdots N} (v_k^{\otimes N})^{\dagger},
\label{simhamhlu}
\ee
where $p_k \equiv t_k/t$ sum up to one.
It already follows from this expression
 that, in general, it is not possible to simulate an arbitrary Hamiltonian $H_{12\cdots N}'$ by $H_{12\cdots N}$ and HLU alone. As an example, suppose that $H_{12\cdots N}$ is symmetric under exchange of qubits. Then $H_{12\cdots N}'$ necessarily has the same symmetry \cite{symmetry}, so that we cannot simulate any inhomogeneous evolution. Thus, the HLU model for simulation is rather restricted. Nevertheless, it is still a useful and ---most important--- available tool to control the evolution of the qubits.

We start our analysis by considering $N=2$ qubits, for which we present most of our results. Later in this section we shall present some generalizations to the general $N$ case.

\subsection{Two-qubit Hamiltonians}

Consider two qubits, $A$ and $B$, that naturally evolve according to some Hamiltonian $H_{AB}$. The Hamiltonians achievable by interspersing the evolution of $H_{AB}$ with control operations $w_k\otimes w_k$, $w_k \in SU(2)$, are now 
\be
H_{AB}^{\prime} = c\sum_{k} p_{k}\ v_{k} \otimes v_{k}\ H_{AB} v_{k}^{\dagger} \otimes v_{k}^{\dagger},  \ \ c\geq 0,
\label{umixing} \ee
where we have followed the same steps than in \cite{IBM}. 
Because the set $\{ I, \sigma_{1}, \sigma_{2}, \sigma_{3} \}$ 
forms a basis for the self-adjoint operators acting on ${\cal C}^{2}$, we can always expand $H_{AB}$ as
\be
H_{AB} = \alpha I \otimes I + h_{A} \otimes I + I \otimes h_{B} + 
  \sum_{i,j=1}^{3} (M)_{ij} \sigma_{i} \otimes \sigma_{j}, 
\label{hamiltonian} \ee
where $M$ is a real $3 \times 3$ matrix called the ``Pauli representation'' of $H_{AB}$, and $h_A$ and $h_B$ are some local terms, that we set to $h_{A}=h_{B} \equiv h$ for simplicity \cite{foot5}.
The term $\alpha I \otimes I$ is an irrelevant phase and we ignore it.
We can always add or eliminate equal local terms by applying an appropriate HLU, 
$e^{-i h t} \otimes e^{-i h t}$, because 
\be
  \left( e^{-i h t}\! \otimes e^{-i h t} \right) e^{-i H_{} t} = 
  e^{-i (h \otimes I + I \otimes h + H_{}) t} + {\mathcal O}(t^{2})\ .
\ee 
Therefore, we can cancel the local terms of $H_{AB}$, simulate the interaction part of $H_{AB}^{\prime}$ using the 
interaction part of $H_{AB}$ and, then, add the local terms of $H_{AB}^{\prime}$ with the corresponding HLU. 
Hence, we need consider Hamiltonians with only interaction terms.
For any unitary $u \in SU(2)$ there always exists a rotation $R \in SO(3)$ such that 
\be 
u \sigma_{i} u^{\dagger} = \sum_{j} R^{T}_{ij} \sigma_{j}. 
\label{so3} \ee 
It is worth mentioning that this argument only works for qubits
\cite{foot2}. 
Using this, we write (\ref{umixing}) as
\be
M^{\prime} = c \sum_{k}\, p_{k}\, R_{k}\, M\, R_{k}^{T}\ .
\label{omixing} \ee
The resulting  $M^{\prime}$ is thus proportional to an orthogonal (real unitary\cite{Bhatia}) mixing of $cM$.
Every term $R_{k} M R_{k}^{T}$ in Eq. (\ref{omixing}) conserves the trace and the symmetry of $M$. 
This implies that
\be
   tr( M^{\prime} )= c \ tr( M ).
\label{jzero}
\ee
Thus, we have no freedom in choosing the time overhead $c$ (inverse of the efficiency $s$ of \cite{IBM}) unless $tr(M)\!=\!tr(M')\!=\!0$.
Another consequence of Eq. (\ref{omixing}), is that with a
 symmetric (antisymmetric) $M$, $M=M^{T}$ ($M=-M^{T}$), we can only 
simulate a symmetric (antisymmetric) $M^{\prime}$. Let us study separately these two cases.

\subsubsection{Symmetric M}



Let us start by considering a Hamiltonian symmetric under the exchange of the subsystems, $S H_{AB} S^\dagger = H_{AB}$, where the swap operator $S$ is defined as to exchange the states of qubit $A$ and $B$, $S (\ket{\psi}_{A} \otimes \ket{\phi}_B) =  \ket{\phi}_A \otimes \ket{\psi}_B$.
We note that this is the case of many relevant interactions, such as the Ising interaction $\sigma_x\otimes \sigma_x$, the $XY$ model or anisotropic exchange interaction $\sigma_x\otimes \sigma_x + \sigma_y\otimes \sigma_y$, and the Heisenberg or isotropic exchange interaction $\sigma_x\otimes \sigma_x + \sigma_y\otimes \sigma_y + \sigma_z\otimes \sigma_z$.

Symmetric real matrices are the only ones that can be diagonalized with $SO(3)$ transformations. 
Calling $\vec{\lambda}$ ($\vec{\lambda}'$) the vector of eigenvalues of $M$ ($M'$) we can rewrite (\ref{omixing}) as 
\be
\lambda'_{i} = c \sum_k\ p_k\ \sum_j (R_k)^2_{ij}\ \lambda_j ,
\label{8} \ee
where we have absorbed the orthogonal matrices that diagonalize $M$ and $M'$ in redefinitions of the $R_k$ matrices.
Doubly stochastic matrices form a convex set with the permutation matrices as the generators (see  Ref. \cite{Bhatia}). 
The squared elements of an orthogonal matrix, $(R)^2_{ij}$, always form a doubly stochastic matrix. Thus, when 
Eq. (\ref{8}) is satisfied there exists a doubly stochastic matrix, $D$, such that 
\be 
\vec{\lambda}' = c\ D \vec{\lambda} .
\label{9} \ee 
On the other hand, for every permutation matrix $P$ there exist orthogonal matrices $R$ fulfilling $(R)^2_{ij}=(P)_{ij}$. 
Therefore, given a doubly stochastic matrix $D$ we always can find an orthogonal mixing such that 
$(D)_{ij} = \sum_k\ p_k\ (R_k)^2_{ij}$.  
Concluding, $M$ simulates $M'$ iff there exists one doubly stochastic matrix $D$ fulfilling Eq. (\ref{9}). The last is 
equivalent to say that $c \vec{\lambda}$ majorizes
\cite{foot3}
 $\vec{\lambda}'$ (the proof is in \cite{Bhatia}).

\vspace{2mm}

{\bf Result 1:} \emph{Let $H_{AB}$ and $H'_{AB}$ be two symmetric interactions. $H_{AB}$ simulates $H_{AB}^{\prime}$ by using fast HLU 
if, and only if, the eigenvalues of $M^{\prime}$ are majorized by the eigenvalues of $c M$, with the appropriate time overhead $c$
given by Eq. (\ref{jzero})}.

\vspace{2mm}

Using the notation of \cite{IBM}, we write 
Result 1 as
\be
  H_{AB}^\prime \leq_{\rm HLU} cH_{AB} \ \  \Longleftrightarrow\ \  \vec{\lambda}^\prime \prec c \vec{\lambda}.
\label{res1}
\ee
Notice that the above simulation strategy is automatically optimal because the
time overhead $c$ is fixed. 

The above result can be understood in physical terms as follows. 
The more isotropic a Hamiltonian is, the less simulating power it
carries. External homogeneous manipulations can only produce mixing, thus isotropy. More precisely, the isotropy of $H_{AB}$ ---as measured by the degree of mixing of $\vec{\lambda}$--- sets the arrow of allowed simulations by HLU. Accordingly,
highly anisotropic Hamiltonians, {\sl e.g.} the Ising model type $H_{AB} = \sigma_z\otimes \sigma_z$, are excellent
resources to simulate other Hamiltonians and, later on, to 
synthesize efficiently all sort of non-local gates. Instead, the isotropic exchange interaction 
$H_{AB} = \sigma_x\otimes\sigma_x+ \sigma_y\otimes\sigma_y + \sigma_z\otimes\sigma_z$ is a fixed point under HLU transformations, 
and can not simulate any other interaction.

\subsubsection{Time--optimal simulation with antisymmetric M}


We consider now the case of an interaction fulfilling $SH_{AB}S^{\dagger} = -H_{AB}$. 
There is an isomorphism between $3\! \times\! 3$ antisymmetric
 matrices and 3-dimensional vectors {\sl via} $v(M)_{i} = \epsilon_{ijk}
 M_{jk}$.
Due to the  invariance of the Levi-Civita tensor $\epsilon_{ijk}$ under rotations, we obtain
 $\vec{v}(R M R^{T})\!=\!R\, \vec{v}(M)$.
Then, we can write Eq.
(\ref{omixing}) in vectorial form:
\be
\vec{v}(M^{\prime}) = c \sum_{k} p_{k}\ R_{k}\ \vec{v}(M).
\label{amixing} \ee
By the triangular inequality we have the following relation among the two moduli 
\be
v(M^{\prime}) \leq c\ v(M). 
\label{cond} \ee

\vspace{2mm}

{\bf Result 2:} \emph{Let $H_{AB}$ and $H'_{AB}$ be two antisymmetric interactions. $H_{AB}$ can always simulate $H^{\prime}_{AB}$ using fast $HLU$ for any time overhead $c$ fulfilling (\ref{cond})}, i.e.,
\be
H_{AB}^\prime \leq_{\rm HLU} c H_{AB} \ \Longleftrightarrow \ v' \leq c\ v.
\ee

\vspace{2mm}

The minimal time overhead or optimal simulation is achieved when the equality (\ref{cond}) holds, $c=v'/v$.
An interesting consequence of Result 2 is that any Hamiltonian with antisymmetric $M$ can simulate itself backwards in time with overhead $1$, also that it can simulate the zero Hamiltonian. In other words, we can always stop its action. We shall use this property later on.


\vspace{2mm}

We finish our analysis for two qubits by noticing that the above cases correspond to the general decomposition of a rank two tensor
in irreducible representations of the rotation group, $J=0$, $J=1$ and
$J=2$. The scalar pieces produces the Eq. (\ref{jzero}) restriction,
the antisymmetric case comes from  $J=1$ and gives Eq. (\ref{amixing}) and
the standard majorization is related to the $J=2$ irreducible representation.

\subsection{Multi-qubit systems with two-body interactions}

Let us now consider a system of $N$ qubits with a Hamiltonian $H_{12\cdots N}$ made out of two-qubit interactions, 
\be
H_{12\cdots N} = \sum_{i,j; j>i} H_{ij}.
\ee
If interactions $H_{ij}$ are arbitrary, the characterization of allowed simulations using $H_{12\cdots N}$ and HLU becomes excessively involved. However, in those systems where, e.g.,  $H_{ij} = H$ for some fixed interaction $H$, we automatically recover the Results 1 and 2 discussed above. This also happens if some couples of qubits do not interact at all. 

As an example, we may consider an array of spins with first-neighbour interaction $\sigma_z\otimes\sigma_z$. By applying convenient HLU, we can make the system evolve according to first-neighbour anisotropic (or isotropic) exchange interactions. More generally, we can achieve any first-neighbour interaction compatible with Eq. (\ref{res1}).


\section{\label{sec3} Time-optimal synthesis of non-local gates using HLU}

The problem of producing a two-qubit gate $\U$ in optimal time using 
a given interaction $H_{AB}$ and fast LU was throughoutly explored in 
\cite{InteractionCost2} and finally solved in
\cite{InteractionCost}. 
In this section we study this problem in the case where 
the given interaction $H_{AB}$ is symmetric ($SH_{AB}S^\dagger = H_{AB}$) and
 only HLU are allowed. Notice that symmetric interactions 
are the most common ones in nature. Because HLU are also symmetric, the net effect of interspersing the action of $H_{AB}$ with 
HLU must give a symmetric $U$. Thus, we only consider unitary gates $U$ fulfilling
\be
  S\,\U S^\dagger = \U .
\label{symmetricU} \ee

\subsection{Synthesis of symmetric two-qubit gates}

In what follows we take advantage of a result of \cite{InteractionCost2,Kraus}, which says that any  
unitary matrix $\U$ acting on the Hilbert space ${\mathcal C}^2 \otimes {\mathcal C}^2$ can be written as 
\be
 \U= u_A \otimes u_B\ \exp{\!\left( -i \sum_{k=1}^3 \lambda'_k\, \sigma_k\! \otimes \sigma_k \right)}\ 
  v_A \otimes v_B ,
\label{Kraus} \ee
the decomposition being unique if we make the restriction 
\be 
\pi/4 \geq \lambda'_1 \geq \lambda'_2 \geq  |\lambda'_3|.
\label{restriction} \ee 
The recipe for getting $\vec{\lambda}'$ is also given in \cite{Kraus}.
When imposing Eq. (\ref{symmetricU}) we get that the one-qubit unitaries $u_A$ and $v_A$ must be equal to $u_B$ and $v_B$ respectively, but since now we cannot modify the sign of the coefficients $\lambda_k$, a unique decomposition for $U$ arises only if we require
 \be 
\pi/4 \geq |\lambda'_1| \geq |\lambda'_2| \geq  |\lambda'_3|.
\label{restriction2} \ee
Because we assume HLU can be performed arbitrarily fast and we are only concerned with simulating times, simulating $\U$ is equivalent to just simulating
\be
  \U_{\vec{\lambda}'} \ \equiv\ \exp{\!\left( -i \sum_{k=1}^3 \lambda'_k\, \sigma_k\! \otimes \sigma_k \right)}.
\ee
The construction $\U_{\vec{\lambda}'}$ is called the canonical form of $\U$.
Note that all commutators $[ \sigma_j \otimes \sigma_j,\, \sigma_k \otimes \sigma_k ]$ vanish and that 
$\exp(-i\pi\, \sigma_k\! \otimes\! \sigma_k /2) = -i \sigma_k\! \otimes\! \sigma_k$ is a HLU. This implies that, for 
any vector $\vec{n} = (n_1, n_2, n_3)$ with integer components $n_k$, 
$\U_{\vec{\lambda}'+ \pi \vec{n}/2} = \U_{\vec{\lambda}'}\, \U_{\pi \vec{n}/2}$, which is equivalent to $\U_{\vec{\lambda}'}$ 
up to HLU. Accordingly, we can associate to $\U$ any one of the Hamiltonians  \cite{permutacio}
\be
  H_{\vec{\lambda}'+ \pi \vec{n}/2} \equiv \sum_{k=1}^3 
  \left(\vec{\lambda}'_k+ \frac{\pi}{2} \vec{n} \right) \sigma_k\! \otimes \sigma_k.
\label{hamiltonians} \ee

Now, if we are able to simulate the action of one of these Hamiltonians for a time $t'=1$, we can implement the 
gate ${\U}$ on our two-qubit system, since $U$ and $U_{\vec{\lambda}'+ \pi \vec{n}/2}$ are HLU--equivalent. In addition, because $H_{AB}$ and $H_{\vec{\lambda}'+ \frac{\pi}{2} \vec{n} }$ are both symmetric, 
Result 1 sets the simulation condition: 
\be
  \exists\ \vec{n}\ \ |\ \ 
  \left( \vec{\lambda}'+ \frac{\pi}{2} \vec{n} \right) \prec c\ \vec{\lambda} .
\label{condition} \ee
That is, even if $H_{AB}$ is not able to simulate $H_{\vec{\lambda}'}$ with HLU, it may still be able to simulate 
$H_{\vec{\lambda}'+\pi\vec{n}/2}$ for some vector $\vec{n}$, and thus perform gate $\U$. Notice, however, that in the case when  
$\vec{\lambda}$ is proportional to $(1,1,1)$ (isotropic interaction) there is no integer vector $\vec{n}$ that makes 
the majorization possible unless $\vec{\lambda}'$ is itself proportional to $(1,1,1)$. In the next paragraph we explore 
for which $\vec{\lambda}$s the condition (\ref{condition}) is satisfied whatever the value of $\vec{\lambda}'$ is.

The sum of the components of $\vec{\lambda}$, $\Sigma_k \lambda_k$, can be positive, 
negative or zero. If $\Sigma_k \lambda_k$ is positive,  
by choosing $n$ large enough, we can make $\vec{\lambda}'+ \frac{\pi}{2}(n,n,n)$ as close as we want to a vector 
proportional to $(1,1,1)$, which is always majorized by any $c\, \vec{\lambda}$ not proportional to $(1,1,1)$.
The same happens if $\Sigma_k \lambda_k$ is negative, but in this case, $n$ must be large and negative. 
If $\Sigma_k \lambda_k=0$, $c\, \vec{\lambda}$ only majorizes vectors having also the sum of their components equal to zero, 
and, not always exists an integer 
vector $\vec{n}$ making $\Sigma_k (\lambda'_k + \pi n_k/2)=0$. We conclude that, interacting hamiltonians with $\Sigma_k \lambda_k$ 
different from zero and $\vec{\lambda}$ not proportional to $(1,1,1)$ can be used to synthesize any symmetric gate.

Notice, from Eq. (\ref{condition}) that increasing the modulus of $\vec{\lambda'}+\pi\vec{n}/2$ by changing $\vec{n}$ carries an 
increase of the time-overhead $c$ (equivalently, decrease of the efficiency $s$ of \cite{IBM}).  
Thus, we can conlcude that the more isotropic $H_{AB}$ is (i.e. the closer $\vec{\lambda}$ is to a vector proportional to $(1,1,1)$)
the less 
efficient $H_{AB}$ is to synthesize gates. As a limiting case, if $\vec{\lambda}$ is
 precisely proportional to $(1,1,1)$, it 
will never majorize any $\vec{\lambda}'$ not proportional to
$(1,1,1)$. 

\subsection{Time-optimal gate synthesis}

So far we have seen that any symmetric gate can be 
synthesized by almost all interactions. A priori,
the best synthesis protocol ought not to make use of a
unique simulated interaction since there is freedom
to change it along the protocol. Yet, the following result
states that, indeed, time-optimal gate synthesis is
achieved maintaining a unique simulated Hamiltonian. As before, $\vec{\lambda}$ and $\vec{\lambda}'$ are the vectors associated to $H_{AB}$ and $\U$, respectively.

\vspace{2mm}

{\bf Result 3 (Theorem 10 in \cite{InteractionCost2}):}
 \emph{The time-optimal way to synthesize a 
symmetric gate $\U$ using the interaction $H_{AB}$ and fast HLU consists
 of simulating, for a time $t'=1$, among all 
Hamiltonians $H_{\vec{\lambda}'+ \frac{\pi}{2} \vec{n} }$ such that
\be 
\, (\vec{\lambda}'+ \frac{\pi}{2} \vec{n} ) \prec c\vec{\lambda}
\ee
 the one with smallest time overhead $c$. 
The minimal interaction time (interaction cost of \cite{InteractionCost}) is given by $c$.}

\vspace{2mm}

The proof of this result follows from particularizing theorem 10 in Ref. \cite{InteractionCost2} as indicated in \cite{Cartan}, when taking into consideration that, up to permutations of the coefficients $\lambda_k$ \cite{permutacio}, Eq. (\ref{hamiltonians}) contains all Hamiltonians $H_{AB}'$ such that $U$ is HLU--equivalent to $\exp(-iH'_{AB})$.

Let us illustrate the above construction with a specific example. We
would like to synthesize the non-local part of the CNOT gate, $\ket{i}_A\ket{j}_B \rightarrow\ket{i}_A\ket{i\oplus j}_B $, by using a physical system which 
 interacts according to the 
 anisotropic exchange interaction: $H= \sigma_x\! \otimes\! \sigma_x + 
\sigma_y\! \otimes\! \sigma_y$, that is $\vec{\lambda}=(1,1,0)$.
We first note that the CNOT gate is equal to:
\bea
  & \U_{\rm CNOT}= \left( e^{i \frac{\pi}{4} \sigma_z}\! \otimes\! U_{\rm Hadamard}e^{i \frac{\pi}{4} \sigma_z} \right) 
  e^{-i\frac{\pi}{4} \sigma_z \otimes \sigma_z} 
\nonumber \\ 
  & \left( I\! \otimes\! U_{\rm Hadamard} \right) \ ,
\label{nonlocalint} \eea 
that is, its interaction part reduces, up to HLU, to
\be
     \label{nonlocalintpiece}
     \U_{(\pi/4,0,0)} = \exp(-i\frac{\pi}{4} \sigma_x\! \otimes\! \sigma_x)\ . 
\ee
A direct simulation of this Hamiltonian cannot be achi\-eved since the
condition $(\pi/4,0,0) \prec c\, (1,1,0)$ cannot be fulfilled for any $c$ (vector $(1,1,0)$ is already more isotropic than $(\pi/4,0,0)$).
Instead, we try to simulate the Hamiltonian 
$H_{(\pi/4,\, \pi/2,\, \pi/2)}$, by which we would also achieve the goal 
according to (\ref{hamiltonians}). 
Now the simulation is possible, 
$(\pi/4,\, \pi/2,\, \pi/2) \prec c\,(1,1,0)$, with time overhead $c=5\pi/8$. 
It can be seen that this is the time-optimal way to do this.
 A possible mixing of permutations of the components of
$\vec{\lambda}=(1,1,0)$ leading to $(\pi/4,\, \pi/2,\, \pi/2)$
 is
\be
  \left( \matrix{  \pi/4\cr \pi/2\cr \pi/2\cr } \right) = 
c \left[  \frac{1}{5} \left( \matrix{  1\cr 1\cr 0\cr } \right) + 
  \frac{1}{5} \left( \matrix{  1\cr 0\cr 1\cr } \right) + 
  \frac{3}{5} \left( \matrix{  0\cr 1\cr 1\cr } \right) \right].  
\ee
Note that  to permute the $x$-$z$($y$-$z$) axes we can perform
 a rotation of $\pi/2$ along the $y$($x$) axis.
 Summing up, in order to synthesize the non-local part of the CNOT gate we have to divide the time $t=5\pi/8$ into small 
equal intervals $\epsilon$ and, in each one of those intervals,
 we have to perform the following sequence:
\bea
  &\left( e^{-i \frac{\pi}{4} \sigma_x} \otimes e^{-i \frac{\pi}{4} \sigma_x} \right)
  e^{-i H \frac{1}{5} \epsilon} 
  \left( e^{i \frac{\pi}{4} \sigma_x} \otimes e^{i \frac{\pi}{4} \sigma_x} \right)  
\nonumber \\ 
  &\left( e^{-i \frac{\pi}{4} \sigma_y} \otimes e^{-i \frac{\pi}{4}
\sigma_y} \right)  
                e^{-i H \frac{1}{5} \epsilon} 
  \left( e^{i \frac{\pi}{4} \sigma_y} \otimes e^{i \frac{\pi}{4} \sigma_y} \right) 
\nonumber \\  
  &e^{-i H \frac{3}{5} \epsilon}
\eea
The gate is synthesized once this protocol is repeated for a time
$t=5\pi/8$.
 A more detailed explanation of the type of manipulations needed above
 is given in \cite{IBM}.


\section{\label{sec4} Hamiltonian simulation and gate synthesis with an Asymmetric interaction and HLU}

Let us now consider how to simulate symmetric interactions and synthesize symmetric gates by using HLU and a two-qubit interaction $H_{AB}$ which is neither symmetric nor antisymmetric under exchange of qubits. 

Recall that we can associate to $H_{AB}$ a real matrix $M$ (its Pauli representation) through Eq. (\ref{hamiltonian}). In what follows we give a method which gets rid of the antisymmetric part of $H_{AB}$, $M_{a} \equiv (M-M^{\rm T})/2$, and the 
local terms, $h_A\!\otimes\!I$ and $I\!\otimes\!h_B$, 
without affecting the symmetric part $M_{s} \equiv (M+M^{\rm T})/2$.
To achieve this goal we use the following property:   
\be
  \sum_{k=0}^3\ \frac{1}{4}\ \sigma_k\! \otimes\! \sigma_k \left( \sigma_n\! \otimes\! \sigma_m \right) 
  \sigma_k^\dagger\! \otimes\! \sigma_k^\dagger = \delta_{nm} \ \sigma_n\! \otimes\! \sigma_n, 
\label{neteja} \ee 
where $n$ and $m$ take the values $0,1,2,3$ and $\sigma_0$ is the identity matrix. 
Performing this unitary mixing on a general Hamiltonian $H$, we get 
\be
  \sum_{k=0}^3\ \frac{1}{4}\ \sigma_k\! \otimes\! \sigma_k\ H\ \sigma_k^\dagger\! \otimes\! \sigma_k^\dagger 
  = \sum_{k=1}^3 (M)_{kk}\, \sigma_k\! \otimes\! \sigma_k. 
\label{neteja2} \ee
That is, only the diagonal part of $M$ survives.
Suppose that $Q$ is the $SO(3)$ matrix that diagonalizes $M_{s}$, and $u$ is the unitary matrix corresponding to $Q$ 
{\sl via} the morphism (\ref{so3}). 
Then, if instead of (\ref{neteja2}), we perform the mixing 
\bea
  & \sum_{k=0}^3\ \frac{1}{4}\ (u^\dagger \sigma_k u) \!\otimes\! (u^\dagger \sigma_k u)\ H\ 
  (u^\dagger \sigma_k u)^\dagger \!\otimes\! (u^\dagger \sigma_k u)^\dagger  \nonumber\\
  & = \sum_{i,j=1}^3 (M_{s})_{ij}\, \sigma_i \!\otimes\! \sigma_j, 
\label{sym} 
\eea
we get the projection onto the symmetric interaction part of the Hamiltonian and no local terms, 
with no loss of efficiency.
In order to make a simulation using only the symmetric interaction part, we divide every 
infinitesimal time  between the action of two HLUs into four equal intervals and perform the mixing (\ref{sym}). 
Therefore, 

\vspace{2mm}

{\bf Result 4:} \emph{Results 1 and 3 also hold in the case when the simulating Hamiltonian $H_{AB}$ is not symmetric.}

\vspace{2mm}

\section{\label{sec5} Universal Quantum computation with an hybrid model of simulation}

We have already argued that an interaction and HLU are not sufficient to perform universal quantum computation, since not all gates can be achieved using only these resources. In this last section we analize which extra resources are needed so that the HLU model recovers universality. As extra resources we consider inhomogeneous LU. A context where this hybrid model may apply is the case of two physical qubits which are brought very close to each other so that they can interact. When the qubits are close, only HLU can be performed in order to control the evolution, and the HLU model applies. But when the systems are brought  far apart and no longer interact, they can be addressed individually and therefore arbitrary LU can be applied.

We have seen in sect. III that any symmetric interaction (a part from the isotropic interaction and interactions with tr$M = 0$) can be used to perform any symmetric gate. This result has been extended in sect. IV also to asymmetric two-qubit interactions (with $M_s$ not proportional to the identity and tr$M \neq 0$). Bearing this in mind, we announce the following result.

\vspace{2mm}

{\bf Result 5:} \emph{Any two-qubit gate can be performed using HLU and any given interaction $H_{AB}$ 
(such that $M_s$ is neither trace-less nor proportional to the identity),
 provided that we allow for two non-homogeneous LU}.

\vspace{2mm}

To justify the last assertion, we recall that any two-qubit gate $U$ can be written as in Eq. (\ref{Kraus}). Thus, we can use the available interaction $H_{AB}$ to perform the non-local part of $U$, which between two non-homogeneous LU interventions leads to $U$. 

This method is very inefficient when $M_s$ is almost proportional to the identity matrix or has very small trace, but alternative methods can be developed in these specific cases, and the hybrid model turns out to be fully universal. For instance, the isotropic interaction $\vec{\lambda}=(1,1,1)$ can be used to perform the gate given by $\vec{\lambda}' = \pi/8(1,1,1)$. Then, we can compose two of these gates with a local unitary to achieve a $(\pi/4,0,0)$ gate, that is the non--local part of a CNOT gate, which together with LU is known to be sufficient for universal quantum computation.

\section{\label{sec6} Conclusions}

We have studied the simulation of non--local Hamiltonians and the synthesis of non-local gates when only HLU can be used to control the available interaction.
This model is motivated by the difficulties that often arise in solid state physics, where individual systems can not be addressed independently because they are very close to each other.

Although notoriously less powerful than inhomogeneous manipulation, homogeneous
local unitary transformations are sufficient to establish simulations within
broad classes of Hamiltonians arranged by their symmetry. In particular, HLU can be used to dispose of the anisotropic terms of any interaction, as required in \cite{Di}.
Symmetric gate synthesis can be  performed using symmetric
interaction Hamiltonians and LHU.
Moreover, these resources can be supplemented with just a few extra non-homogeneous
transformations to fully simulate any two-qubit unitary
transformation, and thereby allow for universal quantum computation.

\bigskip

We thank E. Jan\'e and A. Ac\'\i n for their useful comments. 
We thank financial support from the following projects: AEN99-0766, 1999SGR-00097, IST-1999-11053. G.V. acknowledges grant HPMF-CT-1999-00200 (Marie Curie fellowship) by the European Community. This work was supported in part by the National Science Foundation 
(of U.S.A.) under Grant No. EIA-0086038.


\begin{references}

\bibitem{NMR} C.P. Slichter, \emph{Principles of Magnetic Resonance} (Sprin\-ger, Berlin, 1996). R.R. Ernst, G. Bodenhausen and A. Wokaun, {\em Principles of Nuclear Magnetic Resonance in One and Two Dimensions} (Oxford University Press, Oxford, 1994).

\bibitem{control} H. Rabitz, R. de Vive-Riedle, M. Motzkus and K. Kompa, Science {\bf 288}, 824 (2000).


\bibitem{Duer} W. D\"ur, G. Vidal, J. I. Cirac, N. Linden and S. Popescu, Phys. Rev. Lett. 87, 137901 (2001).

\bibitem{Dodd} J.L. Dodd, M.A. Nielsen, M.J. Bremner and R.T. Thew, 
\emph{Universal quantum computation and simulation using any entangling Hamiltonian and local unitaries}, quant-ph/0106064.

\bibitem{Wocjan} P. Wocjan, D. Janzing and Th. Beth, 
\emph{Simulating Arbitrary Pair-Interactions by a Given Hamiltonian: Graph-Theoretical Bounds on the Time Complexity},
quant-ph/0106077.

\bibitem{IBM} C. H. Bennett, J. I. Cirac, M. S. Leifer, D. W. Leung,
N. Linden, S. Popescu and G. Vidal, 
\emph{Optimal simulation of two-qubit Hamiltonians using general local operations}; quant-ph/0107035.

\bibitem{Bremner} M. A. Nielsen, M. J. Bremner and J. L. Dodd, \emph{Universal simulation of Hamiltonian dynamics for qudits}; 
quant-ph/0109064.

\bibitem{foot1}Operations that can be applied in times which are short compared to  the typical time scale $\tau_{H}$ associated to $H$, $\tau_{H}=(e_{max}-e_{min})^{-1}$, where $e_{max}$ ($e_{min}$) are the highest (lower) eigenvalue of $H$.

\bibitem{InteractionCost2} N. Khaneja, R. Brockett and S. J. Glaser, \emph{Time optimal control in spin systems};
Phys. Rev. A, Vol. 63,  032308 (2001).

\bibitem{InteractionCost} G. Vidal, K. Hammerer and J. I. Cirac,
\emph{Interaction cost of non-local gates}; quant-ph/0112168.

\bibitem{Fort} Fortschritte der Physik {\bf 48} 9-11 pp 769-1138 (2000).

\bibitem{Di} E. Bacon, J. Kempe, D. A. Lidar and K. B. Whaley, {\em Universal 
Fault-Tolerant Quantum Computation on Decoherence-Free Subspaces}; Phys. Rev. 
Lett. 85, 1758 (2000).

D. Bacon, J. Kempe, D. P. Di Vincenzo, D. A. Lidar and K. B. Whaley, 
{\em Encoded Universality in Physical Implementations of a Quantum Computer}; 
quant-ph/0102140

D. P. Di Vincenzo, D. Bacon, J. Kempe, G. Burkard and K. B. Whaley, {\em Universal Quantum Computation with the Exchange Interaction}; quant-ph/0005116

\bibitem{Di2} N.E. Bonesteel, D. Stepanenko and D.P. DiVincenzo, {\em Anisotropic Spin Exchange in Pulsed Quantum Gates}; quant-ph/0106161.

\bibitem{symmetry} Notice that this has important consequences for information processing, since irreducible representations of the permutation group are left invariant under symmetric evolutions, and the dimension of the largest representation grows only linearly with $N$. Therefore in this highly symmetric setting we cannot even exploit the exponential growth (in $N$) of dimension offered by quantum computers, and the system becomes useless for quantum computation unless some degree of inhomogeneity can be introduced.

\bibitem{QC} M. A. Nielsen and I. L. Chuang, \emph{Quantum computation and quantum information}; Cambridge University Press, Cambridge, U.K., 2000.




\bibitem{foot5} This is equivalent to assume that each qubit is equally affected by the environment.

\bibitem{foot2}For qunits this expression works similarly 
with $U\in SU(n)$ and $R\in SO(n^2-1)$. Nevertheless the counting of degrees of freedom shows that this
$R$ belongs to a subset of $SO(n^2-1)$ and, in general, will not allow
for a diagonalization of $M$ unless this latter matrix is severely restricted. 
A complete study of hamiltonian simulation for qunits is made in \cite{Bremner}.
 
\bibitem{Bhatia} Rajendra Bhatia \emph{Matrix Analysis}, Chapter II.

M. A. Nielsen and G. Vidal, \emph{Majorization and the interconversion of bipartite states};
Quantum Information and Computation, Vol. 1,  1 (2001) 76.

\bibitem{foot3}We say that a real vector $\vec{x}$ is majorized by a real 
vector $\vec{y}$, written $\vec{x} \prec \vec{y}$, iff
\bea 
  & x^{\downarrow}_{1} \leq y^{\downarrow}_{1} 
\nonumber \\ 
  & x^{\downarrow}_{1}+x^{\downarrow}_{2} \leq y^{\downarrow}_{1}+y^{\downarrow}_{2} 
\nonumber \\  
  & x^{\downarrow}_{1}+x^{\downarrow}_{2}+x^{\downarrow}_{3}=y^{\downarrow}_{1}+y^{\downarrow}_{2}+y^{\downarrow}_{3} \ \ , 
\nonumber \eea
where $x^{\downarrow}_{i}$ is the $i^{\rm th}$ component of $\vec{x}$
when we rearrange them in decreasing order. 

\bibitem{Kraus} B. Kraus and J. I. Cirac, \emph{Optimal Creation of Entanglement Using a Two-Qubit Gate}; 
quant-ph/0011050. 

\bibitem{permutacio} Notice that by means of HLU we can permute the $x,y,z$ axes of both qubits $A$ and $B$ simultaneously. For instance, the transformation $u\otimes u$, $u = (I-i\sigma_x)/\sqrt{2}$, permutes axes $y\leftrightarrow z$ of the two qubits. Thus, an interaction $H_{AB}$ with vector $\vec{\lambda}=(\lambda_x,\lambda_y,\lambda_z)$ can simulate any interaction whose vector components are permutations of $(\lambda_x,\lambda_y,\lambda_z)$. Also, in gate synthesis, a gate $U$ with vector $\vec{\lambda}'=(\lambda_x',\lambda_y',\lambda_z')$ is HLU--equivalent to a gate with permuted components. The majorization relation $\prec$ is already defined to be invarinat under permutations, and this is why we do not consider permuted versions of $vec{\lambda'}$ in Eq. (\ref{hamiltonians}).

\bibitem{Cartan} Theorem 10 of \cite{InteractionCost2} holds for any Cartan decomposition of a Lie algebra. 
In our particular case, the Cartan decomposition is
\bea 
  & g = k \oplus p 
\nonumber \\
  & k \equiv \{ I\!\otimes\!\sigma_i + \sigma_i\!\otimes\!I: i=1,2,3 \} 
\nonumber \\ 
  & p \equiv \{ \sigma_i\!\otimes\!\sigma_i, \sigma_i\!\otimes\!\sigma_j + \sigma_j\!\otimes\!\sigma_i:  i,j=1,2,3 \} \ \ . 
\nonumber \eea  
 
\end{references}
\end{document}